\documentclass[twocolumn,showpacs,preprintnumbers,amsmath,amssymb]{revtex4}

\usepackage{graphicx}
\usepackage{dcolumn}
\usepackage{bm}

\begin{document}
\title{A microfabricated surface-electrode ion trap in silicon}
\preprint{quant-ph/0605170}

\author{J.~Britton}
\email{britton@nist.gov}
\author{D.~Leibfried}
\author{J.~Beall}
\altaffiliation{NIST, Quantum Electrical Metrology Division,
Boulder, CO}
\author{R.~B.~Blakestad}
\author{J.~J.~Bollinger}
\author{J.~Chiaverini}
\altaffiliation[Present Address: ]{Los Alamos National Laboratory}
\author{R.~J.~Epstein}
\author{J.~D.~Jost}
\author{D.~Kielpinski}
\altaffiliation[Present Address: ]{Griffith University, Brisbane,
Australia}
\author{C.~Langer}
\author{R.~Ozeri}
\author{R.~Reichle}
\altaffiliation[Present Address: ]{University of Ulm, Germany}
\author{S.~Seidelin}
\author{N.~Shiga}
\author{J.~H.~Wesenberg}
\author{D.~J.~Wineland}
\affiliation{NIST, Time and Frequency Division, Boulder, CO}

\date{\today}

\begin{abstract}
The prospect of building a quantum information processor underlies
many recent advances ion trap fabrication techniques.  Potentially,
a quantum computer could be constructed from a large array of
interconnected ion traps.  We report on a micrometer-scale ion trap,
fabricated from bulk silicon using micro-electromechanical systems
(MEMS) techniques. The trap geometry is relatively simple in that
the electrodes lie in a single plane beneath the ions.  In such a
trap we confine laser-cooled $^{24}$Mg$^+$ ions approximately 40
$\mu$m above the surface.  The fabrication technique and planar
electrode geometry together make this approach amenable to scaling
up to large trap arrays. In addition we observe that little laser
cooling light is scattered by the electrodes.

\end{abstract}

\maketitle

Trapped ions are an attractive system for large scale quantum
information processing (QIP) \cite{cirac2000,arda_roadmap}. All the
building blocks of such a processor are demonstrated in the
laboratory, and there is no known fundamental obstacle to scaling up
to a large processor. Nevertheless considerable challenges remain,
including methods of manipulating large arrays of ion qubits. Some
proposed schemes for QIP in ions involve many interconnected ion
traps with integrated optics and control electronics
\nocite{devoe1998,bible,cirac2000,kielpinski02,duan04,kim05}
\cite{devoe1998,bible,cirac2000,kielpinski02,duan04,kim05}. This
paper describes a new approach for making trap structures that is
relatively straightforward and holds the potential of scaling up.

\begin{figure}[t]
\centering\includegraphics[width=8.6cm]{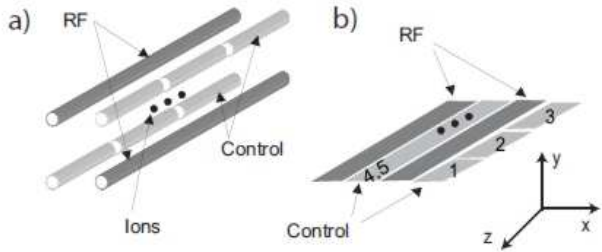}
\caption{\label{fig:schematicview}(a) An idealized geometry for a
linear RF Paul trap; (b) A surface-electrode geometry where all
electrodes reside in a single plane, with the ions trapped above
this plane. \cite{chiaverini2005b}}
\end{figure}

A straightforward implementation of a linear radio-frequency (RF)
ion trap consists of four parallel conducting rods as depicted in
Figure \ref{fig:schematicview}(a) \cite{paul90}. With opposite pairs
of rods held at RF-ground and RF potential respectively, a
ponderomotive potential arises that provides nearly harmonic
confinement in the $\hat{x}$-$\hat{y}$ radial plane, as in a RF
quadrupole mass filter. The RF-grounded rods can be segmented (as
illustrated) and a static positive potential (for positive ions) is
applied to the end segments relative to the inner segments
\cite{raizen92}. This defines a trap zone for charged particles in
the $\hat{z}$ axial direction. This type of trap has been used in
ion QIP work but may be difficult to implement on a large scale. Key
trap characteristics for large arrays of qubits include a small
ion-electrode spacing, accurately defined smooth electrode surfaces,
low RF losses, compatibility with on-chip control electronics, a low
anomalous heating rate
\nocite{turchette2000a,bible,chiaverini2005b,deslauriers06}
\cite{turchette2000a,bible,chiaverini2005b,deslauriers06}, and the
possibility of scaling to many trapping zones
\cite{arda2004solicitation,kim05}. No current trap technology meets
all these requirements.

\begin{figure}[t]
\centering\includegraphics[width=8.6cm]{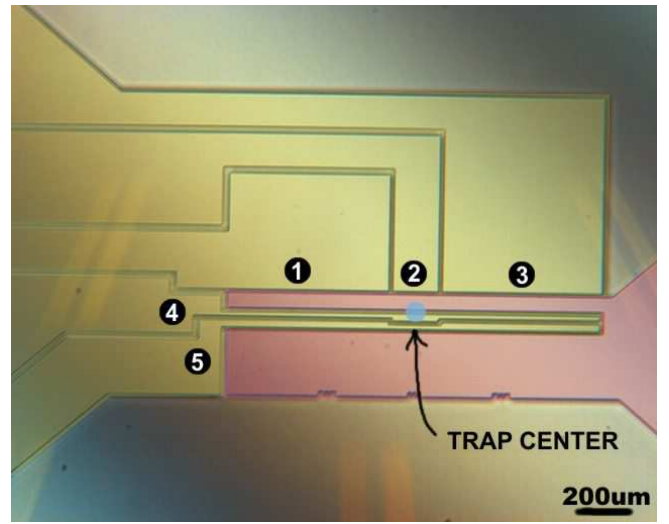}
\caption{\label{fig:optical}An image from an optical microscope
showing the boron-doped silicon trap viewed normal to the trap
surface. False coloring emphasizes the RF electrodes (pink) and
control electrodes (yellow).  The control electrodes are numbered.
Alignment marks on the edge of the outermost RF electrode assist
with laser beam alignment.  The images in Figs. \ref{fig:threeions}
and \ref{fig:twelveions} were taken in the same view as this
figure.}
\end{figure}

Among several RF ion-trap electrode configurations
\nocite{schrama93,raizen92,ghosh1995,rowe02,blain04,home04,barrett04,wineland05,hensinger06,stick2006,seidelin06,brown2006}
\cite{schrama93,raizen92,ghosh1995,rowe02,blain04,home04,barrett04,wineland05,hensinger06,stick2006,seidelin06,brown2006},
two approaches have emerged as possible candidates for scaling to
large arrays of traps. One approach uses multilayered structures
with large trap depths and open optical access but which are
relatively difficulty to fabricate because of the number of layers
\nocite{rowe02,home04,barrett04,wineland05,hensinger06,stick2006}
\cite{rowe02,home04,barrett04,wineland05,hensinger06,stick2006}.
Two microfabricated multilayer traps have been demonstrated: in
boron-doped silicon \cite{wineland05} and in
aluminum-gallium-arsenide \cite{stick2006}. Alternately,
single-layer traps, wherein all electrodes lie in a single plane,
are more amenable to microfabrication and integration with on-chip
control electronics \cite{chiaverini2005b,kim05}, see Figure
\ref{fig:schematicview}b. Recently this approach was demonstrated
for atomic ions in a trap with gold electrodes on a fused silica
substrate \cite{seidelin06} and with copper electrodes on a
GML-1000 printed circuit board \cite{brown2006}.

\begin{figure}[t]
\centering\includegraphics[width=6.5cm]{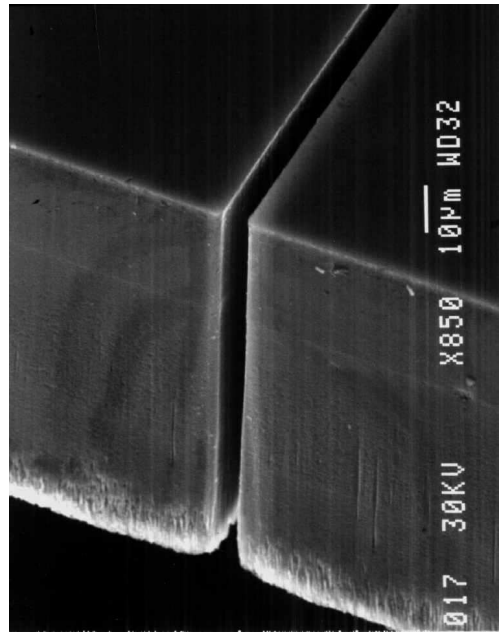}
\caption{\label{fig:smooth}A SEM image of a $100$ $\mu$m thick
silicon wafer after through-wafer etching.  The dark surface near
the top is the wafer's polished top side and was untouched by the
etch. This surface was coated with photoresist, then
photolithographically patterned to expose bare silicon on the entire
lower side and along a $5$ $\mu$m wide strip. The wafer was then
subjected to etching. In an actual trap the ions would lie above the
top surface.  The perspective in this image emphasizes (a) the
accuracy of the features, (b) the high aspect ratio of the
through-wafer etch, and (c) the relative smoothness of the etch
sidewalls. The ragged edge at the bottom of the etch is an artifact
and can be minimized with proper care. }
\end{figure}

\begin{figure}[t]
\centering\includegraphics[width=8.6cm]{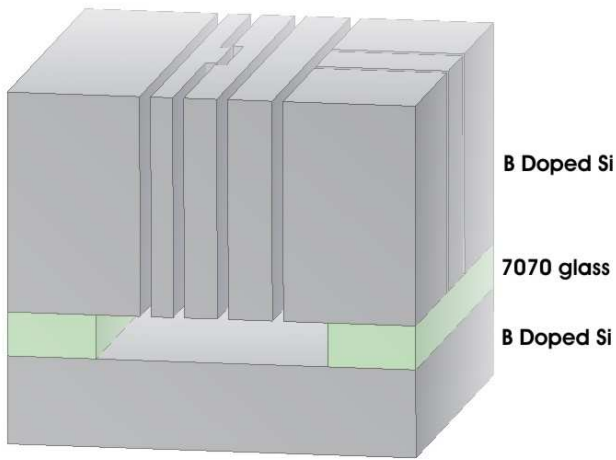}
\caption{\label{fig:slice}A schematic view of the trap. A trapped
ion would lie above the uppermost layer of silicon. The silicon
electrodes overhang the glass frame. In the real structure the glass
lies several millimeters away from the trap region.}
\end{figure}

Here we report on the demonstration of a different fabrication
technique for surface-electrode traps where the electrodes are
lithographically defined from the bulk of a silicon wafer. While the
geometry is similar to that in \cite{seidelin06}, the fabrication
method described here has several advantages: the process uses only
silicon, so it is compatible with CMOS foundry processes
\cite{kim05}; surfaces exposed to the ion have low surface roughness
($<1$ nm RMS); and stray fields from insulators are minimized since
the trapping region lies far from the nearest insulating surface,
see Figures \ref{fig:optical},\ref{fig:smooth}, and \ref{fig:slice}.
Using this technology, multizone traps with electrodes $10$ $\mu$m
wide are practical.

A demonstration trap was fabricated by through-wafer etching of a
commercially available boron doped silicon wafer.  The electrode
pattern was defined by standard photolithography, then etched using
deep reactive ion etching (DRIE), a standard MEMS technique also
known as the Bosch process
\nocite{bosch1995,madou1997}\cite{bosch1995,madou1997}. This etch
can produce trenches of arbitrary lateral geometry, hundreds of
micrometers deep and as small as several micrometers wide while
maintaining an etch aspect ratio of $100:1$ and low sidewall
roughness, see Figure \ref{fig:slice}. The commercially available
silicon is doped with boron as the silicon crystal is grown yielding
resistivities of $\rho \simeq 500\times 10^{-6}$ ohm-cm; no
additional metalization is needed in the trap region. For reference,
at room temperature $\rho($Au$) \simeq 2\times 10^{-6}$ ohm-cm,
$\rho($Ti$)\simeq 40\times 10^{-6}$ ohm-cm and $\rho ($W$) \simeq
5\times 10^{-6}$ ohm-cm \cite{crc2006}.

After etching, structural support is provided by anodic bonding of
the silicon wafer to a glass substrate
\nocite{tong1998,kielpinski2001}\cite{tong1998,kielpinski2001}. The
glass has a gap beneath the trapping region above which the silicon
electrodes are cantilevered in vacuum, see Figure \ref{fig:slice}.
This gap is present to keep dielectric surfaces away from the trap
region and to reduce the penetration of RF fields into the glass.
The glass is a borosilicate ($7070$) with a loss tangent of $0.060$
($1$ MHz, $20^\circ$ C) and a coefficient of thermal expansion (CTE)
matched to silicon. For reference, the loss tangent of Pyrex
($7740$) is $0.500$ and that of $99.5\%$ pure alumina is $0.0001$
\cite{corning7070}.

The anodic bonding is done after thorough cleaning to remove organic
contaminants and the native silicon oxide by use of solutions of
$8$:$1$ :: H$_2$SO$_4$ : H$_2$O$_2$ at $90^\circ$ C and $6$:$1$ ::
$40$\% NH$_4$F : $49$\%HF at $25^\circ$ C, respectively. The etch
geometry is arranged so that all the electrodes are joined at the
periphery until after bonding, whereupon they are separated by a
perpendicular cut with a dicing saw \cite{kielpinski2001}.  For
wiring to external leads, contact pads are made by evaporative
deposition of Ti and Cu near the periphery of the chip.

The electrode geometry used here is shown in Figure
\ref{fig:optical}. Five independent static-potential control
electrodes are present in this design. The control electrodes define
the axial potential curvature and provide sufficient degrees of
freedom to properly overlap the static electric field with the RF
pseudopotential minimum, that minimizes ion micromotion in both
radial directions \cite{chiaverini2005b,seidelin06}.

Capacitive coupling of the RF and control electrodes due to their
close proximity can induce an undesirable RF potential on the
latter; this is minimized by coupling the control electrodes to
ground through low-pass RC filters mounted on a ceramic chip carrier
within the vacuum housing \cite{rowe02}.

Ions are created by laser photoionization (PI) of an atomic vapor
evolved from a resistively heated ampule containing $^{24}$Mg. In
some trap designs there is risk of this vapor causing shorts
between trap electrodes \cite{pearson2005,seidelin06,brown2006};
the cantilevered geometry avoids this problem as vacuum separates
the electrodes in the trapping region. For PI a laser is tuned to
the 3s$^{2~1}$S$_0$~$\leftrightarrow$~3s3p~$^{1}$P$_1$ electric
dipole transition in neutral magnesium at $285$ nm
\nocite{madsen2000,madsen2002}\cite{madsen2000,madsen2002}. From
the 3s3p~$^{1}$P$_1$ state, absorption of another photon at $285$
nm can excite the electron to the continuum. The PI beam is
present in the trap region only during the loading phase.

The ions are Doppler cooled with a $280$ nm laser tuned $400$ MHz
red of the 3s~$^{2}$S$_{1/2}$~$\leftrightarrow$ 3p $^{2}$P$_{1/2}~$
electric-dipole transition in $^{24}$Mg$^+$. The laser beam
propagates parallel to the trap surface, at an angle of $45^\circ$
with respect to the trap $\hat{z}$-axis, see Figures
\ref{fig:optical} and \ref{fig:schematicview}b.  The radial trap
axes lie at approximately $45^\circ$ with respect to the trap
surface. Therefore, the laser has a projection along all three axes,
allowing laser cooling along all axes \cite{chiaverini2005b}.

The cooling and PI beams copropagate. They have a power of $500$
$\mu$W and $800$ $\mu$W respectively, measured after they exit the
vacuum apparatus. At the trapping region they have a waist of
approximately $40$ $\mu$m and are centered approximately $40$ $\mu$m
above the surface. Ions are detected by observing the ion
fluorescence on a CCD camera. The background includes stray light
scattered from the cooling laser beam  by the apparatus. The
signal-to-background for a single ion is higher than $100:1$ even
with a beam intensity $40$ times the resonance saturation intensity,
limited by noise due to the CCD chip.

\begin{figure}[t]
\centering\includegraphics[width=4cm]{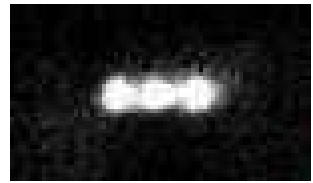}
\caption{\label{fig:threeions}A linear crystal of three ions along
the trap $\hat{z}$-axis. With potentials as described in the text
the length of the crystal is $13.6$ $\mu$m, as expected for the
measured axial frequency of $760$ kHz.}
\end{figure}

Initial trap parameters (RF and control electrode potentials) were
determined numerically by the boundary element method. The primary
solution constraint is that the micromotion be nulled as discussed
above.  An ion was first trapped with static potentials of $V1 =
0.32$, $V2 = 0.72$, $V3 = 0.74$, $V4 = -0.90$ and $V5 = 1.00$ volts
and estimated peak RF potential of $125$ $V$ amplitude with respect
to DC ground at $87$ MHz. For these parameters the expected trap
depth is approximately $200$ meV, and the ions are located $40$
$\mu$m above the surface. Figure \ref{fig:threeions} shows a linear
crystal of three ions trapped using these potentials.

\begin{figure}[t]
\centering\includegraphics[width=8.6cm]{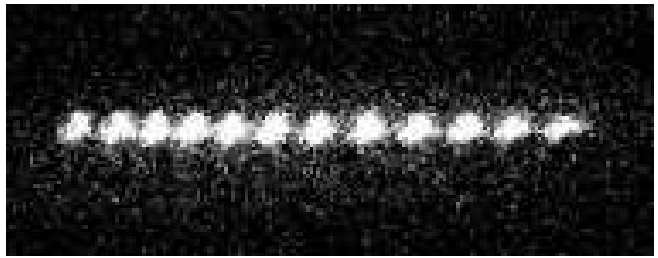}
\caption{\label{fig:twelveions}A linear crystal of 12 ions along
trap $\hat{z}$-axis. The potential on the endcaps is $V1 = -800$ mV
and  $V3 = -600$ mV. The length of the crystal in the image is $103$
$\mu$m. }
\end{figure}

The axial oscillation frequency was measured experimentally by
applying oscillating potentials to electrode $1$. When resonant with
a motional mode of the ions, the motion is excited.  We detect this
by observing a change in the ion's fluorescence due to Doppler
broadening \cite{jefferts95}. With the trap parameters listed above,
for three ions an axial frequency of $760$ kHz was measured. Figure
\ref{fig:twelveions} shows a linear crystal of twelve ions achieved
by lowering the axial potential on the endcaps to $V1 = -800$ mV,
$V2 = 0$ mV and $V3 = -600$ mV.  This configuration holds the ions
in a linear chain due to a relatively strong axial component of the
RF pseudopotential.  Single ion lifetimes greater than one hour were
observed.

Heating of the ion motion is a concern in all ion traps and is an
important consideration in evaluating new fabrication techniques
\nocite{monroe1995,bible,turchette2000a,deslauriers06}\cite{monroe1995,bible,turchette2000a,deslauriers06}
for QIP. Therefore, the next steps in the evaluation of this trap
will include characterization of the heating at the quantum level.

It is important to investigate a variety of approaches to ion trap
fabrication on the way towards larger QIP processors.  Particular
attention must be paid to the practicality of scaling to many
trapping zones and to material science issues raised by anomalous
heating.

This work was supported by the Advanced Research and Development
Activity (ARDA) under contract $MOD-7171.05$ and NIST.  S.~S. thanks
the Carlsberg Foundation for financial support; J.~H.~W. thanks the
Danish Research Agency.  This manuscript is a publication of NIST
and not subject to U.S. copyright.

\newpage 
\bibliography{siflattrap}

\end{document}